\documentclass[aps,pra,reprint,superscriptaddress]{revtex4-1}

\usepackage{graphicx}
\usepackage{bm}
\usepackage{bbold}

\renewcommand{\vec}[1]{{\mathbf #1}}

\begin{document}

\title{Control of light trapping in a large atomic system by a static magnetic field}

\author{S.E. Skipetrov}
\email[]{Sergey.Skipetrov@lpmmc.cnrs.fr}
\affiliation{Universit\'{e} Grenoble Alpes, LPMMC, F-38000 Grenoble, France}
\affiliation{CNRS, LPMMC, F-38000 Grenoble, France}

\author{I.M. Sokolov}
\email[]{ims@is12093.spb.edu}
\affiliation{Department of Theoretical Physics, Peter the Great St. Petersburg Polytechnic University, 195251 St. Petersburg, Russia}

\author{M.D. Havey}
%% \email[]{...}
\affiliation{Department of Physics, Old Dominion University, Norfolk, VA, 23529 USA}

\date{\today}

\begin{abstract}
We propose to control light trapping in a large ensemble of cold atoms by an external, static magnetic field. For an appropriate choice of frequency and polarization of the exciting pulse, the field is expected to speed up the fluorescence of a dilute atomic system. In a dense ensemble, the field does not affect the early-time superradiant signal but amplifies intensity fluctuations at intermediate times and induces a very slow, nonexponential long-time decay. The slowing down of fluorescence is due to the excitation of spatially localized collective atomic states that appear only under a strong magnetic field and have exponentially long lifetimes. Our results therefore pave a way towards experimental observation of the disorder-induced localization of light in cold atomic systems.
\end{abstract}

\maketitle

\section{Introduction}
\label{sec:intro}

Manipulating quantum states of atomic ensembles opens interesting possibilities for storage and, potentially, processing of quantum information \cite{divincenzo2000, monroe2002}. Particularly interesting are long-lived atomic states that can be used to trap an optical excitation (a photon) for a time $\tau$ much exceeding the lifetime $\tau_0$ of the excited state of an isolated atom. A very efficient way to trap a photon in an atomic medium is based on the use of the phenomenon of electromagnetically induced transparency (EIT) \cite{lukin2003,flei2005} that is accompanied by a very slow or even vanishing group velocity of a weak probe pulse in a medium manipulated by a strong control beam \cite{phillips2001,liu2001}. Another phenomenon that can substantially delay a photon in an atomic medium is Dicke subradiance \cite{dicke54, gross82}. In contrast to EIT, it is a linear, single-photon effect relying on the interference of electromagnetic emissions of different atoms adding up destructively and leading to a slowing down of the decay of a collective atomic state with time. Recent theoretical \cite{bienaime12, scully15} and experimental \cite{guerin16} results suggest that the lifetime $\tau$ of a subradiant state should grow roughly linearly with the size $R$ of the atomic system. A larger $\tau \propto R^2$ can be obtained under conditions of diffuse scattering of near-resonant light in relatively dense atomic clouds \cite{labeyrie03}. Here we propose and theoretically investigate a different mechanism for trapping of light in an atomic system---a mechanism making use of a disorder-induced spatial localization of collective atomic states \cite{anderson58, skip15}. The lifetimes of the latter grow {\em exponentially} with the sample size $R$ and under realistic conditions, can be many orders of magnitude longer than the lifetimes of both the subradiant states and the states that build up in the regime of diffuse light scattering. There is a price to pay, however, since the disorder-induced localization of light in three dimensions turns out to be difficult to achieve \cite{sperling16,skip16} and was predicted to be even impossible in an atomic ensemble in the absence of external fields \cite{skip14}. Localized states may appear only in dense atomic systems under sufficiently strong magnetic fields \cite{skip15}. This additional complication can be turned into advantage because the field can be used to tune the spatial localization of atomic states and their decay with time thus providing an efficient means of control over the trapping of light.

\section{The model}
\label{sec:model}

We consider an experimentally relevant situation of a short laser pulse (central frequency $\omega_{\mathrm{L}}$, duration $\tau_{\mathrm{L}}$) incident on a spherical cloud (radius $R$, volume $V$) of $N$ two-level atoms located at random positions $\mathbf{r}_i \in V$ ($i = 1, \ldots, N$). An atom $i$ has a ground state $|\mathrm{g}_i \rangle$ with the total angular momentum $J_{\mathrm{g}} = 0$, an excited state $|\mathrm{e}_i \rangle$ with $J_{\mathrm{e}} = 1$, a transition frequency $\omega_0$, and a natural lifetime of the excited state $\tau_0 = 1/\Gamma_0$. The atoms are subject to a static, spatially uniform magnetic field $\mathbf{B}$ that splits the otherwise triply degenerate excited state $|\mathrm{e}_{i} \rangle$ into three substates $|\mathrm{e}_{im} \rangle$ corresponding to the magnetic quantum numbers $m = 0$, $\pm 1$, respectively. The Hamiltonian of the atomic system interacting with the free electromagnetic field is \cite{sigwarth13,skip15}
\begin{eqnarray}
{\hat H} &=& \sum\limits_{i=1}^{N} \sum\limits_{m=-1}^{1} \left(
\hbar \omega_0 + g_{\mathrm{e}} \mu_{\mathrm{B}} B m \right) | \mathrm{e}_{im} \rangle
\langle \mathrm{e}_{im}|
\nonumber \\
&+&
\sum\limits_{\mathbf{s} \perp \mathbf{k}} \hbar ck
\left( {\hat a}_{\mathbf{k} \mathbf{s}}^{\dagger} {\hat a}_{\mathbf{k}\mathbf{s}} + \frac12 \right)
- \sum\limits_{i=1}^{N} {\hat \mathbf{D}}_i \cdot {\hat \mathbf{E}}(\mathbf{r}_i)
\nonumber \\
&+& \frac{1}{2 \epsilon_0}
\sum\limits_{i \ne j}^{N} {\hat \mathbf{D}}_i \cdot {\hat \mathbf{D}}_j \delta(\mathbf{r}_i - \mathbf{r}_j),
\label{ham}
\end{eqnarray}
where ${\hat \mathbf{D}}_i$ are the atomic dipole operators, ${\hat \mathbf{E}}(\mathbf{r}_i)$ is the electric displacement vector divided by the vacuum permittivity $\epsilon_0$, ${\hat a}_{\mathbf{k} \mathbf{s}}^{\dagger}$ and ${\hat a}_{\mathbf{k}\mathbf{s}}$ are the photon creation and annihilation operators corresponding to a mode of the free electromagnetic field having a wave vector  $\vec{k}$ and a polarization $\vec{s}$, $2\pi\hbar$ is the Planck's constant, $\mu_{\mathrm{B}}$ is the Bohr magneton, and $g_{\mathrm{e}}$ is the Land\'{e} factor of the excited state.

As we have shown previously \cite{skip14,skip15}, it is convenient to introduce a $3N \times 3N$ Green's matrix $G$ of the considered random configuration of atoms:
\begin{eqnarray}
G_{\mathrm{e}_{i m} \mathrm{e}_{j m'}} &=& \left(i -  {2 m \Delta} \right) \delta_{\mathrm{e}_{i m} \mathrm{e}_{j m'}} -
\frac{2}{\hbar \Gamma_0} (1 - \delta_{\mathrm{e}_{i m} \mathrm{e}_{j m'}})
\nonumber \\
&\times&
\sum\limits_{\mu, \nu}
{d}_{\mathrm{e}_{i m} \mathrm{g}_i}^{\mu} {d}_{\mathrm{g}_j \mathrm{e}_{j m'}}^{\nu}
\frac{e^{i k_0 r_{ij}}}{r_{ij}^3}
\nonumber
\\
&\times& \left\{
\vphantom{\frac{r_{ij}^{\mu} r_{ij}^{\nu}}{r_{ij}^2}}
 \delta_{\mu \nu}
\left[ 1 - i k_0 r_{ij} - (k_0 r_{ij})^2 \right]
\right.
\nonumber \\
&-&\left. \frac{r_{ij}^{\mu} r_{ij}^{\nu}}{r_{ij}^2}
\left[3 - 3 i k_0 r_{ij} - (k_0 r_{ij})^2 \right]
\right\}.
\label{green}
\end{eqnarray}
Here $k_0 = \omega_0/c$, $\Delta = g_{\mathrm{e}} \mu_{\mathrm{B}} B/\hbar\Gamma_0$ is the Zeeman shift in units of $\Gamma_0$, $\vec{d}_{\mathrm{e}_{i m} \mathrm{g}_i} = \langle J_{\mathrm{e}} m|{\hat \mathbf{D}}_i | J_{\mathrm{g}} 0 \rangle$, and
$\vec{r}_{ij} = \vec{r}_i - \vec{r}_j$.
The state of the atomic system can be represented in terms of eigenvectors $\mathbf{\Psi}_n$ of $G$ to which we will therefore refer as `quasi-modes'. The eigenvalues $\Lambda_n$ of $G$ yield the frequencies $\omega_n = \omega_0 - (\Gamma_0/2) \mathrm{Re} \Lambda_n$ and decay rates $\Gamma_n/2 = (\Gamma_0/2) \mathrm{Im} \Lambda_n$ of quasi-modes.
The Green's matrix defines a projection of the resolvent operator on the eigenstates of the Hamiltonian (\ref{ham}) with one excited atom and no photons \cite{sokolov11}:
\begin{eqnarray}
{\cal R}(\omega) = \left[(\omega-\omega_0) \mathbb{1} + (\Gamma_0/2) G \right]^{-1},
\label{resolvent}
\end{eqnarray}
where $\mathbb{1}$ is a $3N \times 3N$ identity matrix. The intensity  $I(\mathbf{\Omega}, t)$ of light that the atoms scatter in a unit solid angle around an arbitrary direction $\mathbf{\Omega} = \{ \theta, \varphi \}$ when illuminated by an incident pulse
$\mathbf{E}(\mathbf{r}, t) = \mathbf{u}_{\mathrm{in}} \int_{-\infty}^{\infty} (d\omega/2\pi) E(\omega)
\exp(i \mathbf{k}_{\mathrm{in}} \mathbf{r} - i \omega t)$ is then \cite{fofanov11}:
\begin{eqnarray}
I(\mathbf{\Omega}, t) &=& \frac{c}{4 \pi \hbar^2}
\left| k_0^2 \int\limits_{-\infty }^{\infty} \frac{d\omega}{2\pi} E(\omega)
\sum\limits_{i,j = 1}^{N}
e^{i (\mathbf{k}_{\mathrm{in}} \mathbf{r}_j - \mathbf{k} \mathbf{r}_i) - i \omega t} \right.
\nonumber \\
&\times& \sum\limits_{\alpha=1}^{2}
\sum\limits_{m, m' =-1}^{1}
(\mathbf{u}'^{*}_\alpha \cdot \mathbf{d}_{\mathrm{g}_i \mathrm{e}_{im}})
{\cal R}_{\mathrm{e}_{im} \mathrm{e}_{jm'}}(\omega )
\nonumber \\
&\times&
\left. (\mathbf{u}_{\mathrm{in}} \cdot \mathbf{d}_{\mathrm{e}_{jm'} \mathrm{g}_j})
\vphantom{\int\limits_{-\infty }^{\infty}}
\right|^{2}.
\label{intensity}
\end{eqnarray}
In spherical coordinates, $\mathbf{k}_{\mathrm{in}} = \{ \omega/c, \mathbf{0} \}$ and $\mathbf{k} = \{ \omega/c, \mathbf{\Omega} \}$ are the wave vectors of incident and scattered waves, respectively. The unit vector $\mathbf{u}_{\mathrm{in}}$ determines the polarization of the incident wave and the vectors $\mathbf{u'_{\alpha}}$ correspond to two possible orthogonal polarization of scattered light. In the following we will use Eq.\ (\ref{intensity}) averaged over the ensemble of possible atomic configurations $\{ \mathbf{r}_i \}$ to study the average intensity of the time-dependent fluorescence $\langle I(\mathbf{\Omega}, t) \rangle$. For illustrative purposes, we will restrict our consideration to rectangular incident pulses having a central frequency $\omega_{\mathrm{L}} = \omega_0 + \delta_{\mathrm{L}} \Gamma_0$, where $\delta_{\mathrm{L}}$ is the detuning of the incident light in units of $\Gamma_0$. However, similar results are expected for any pulse of well-defined duration.

\section{Long-time fluorescence}
\label{sec:long}

\begin{figure}
\includegraphics[width=0.9\columnwidth]{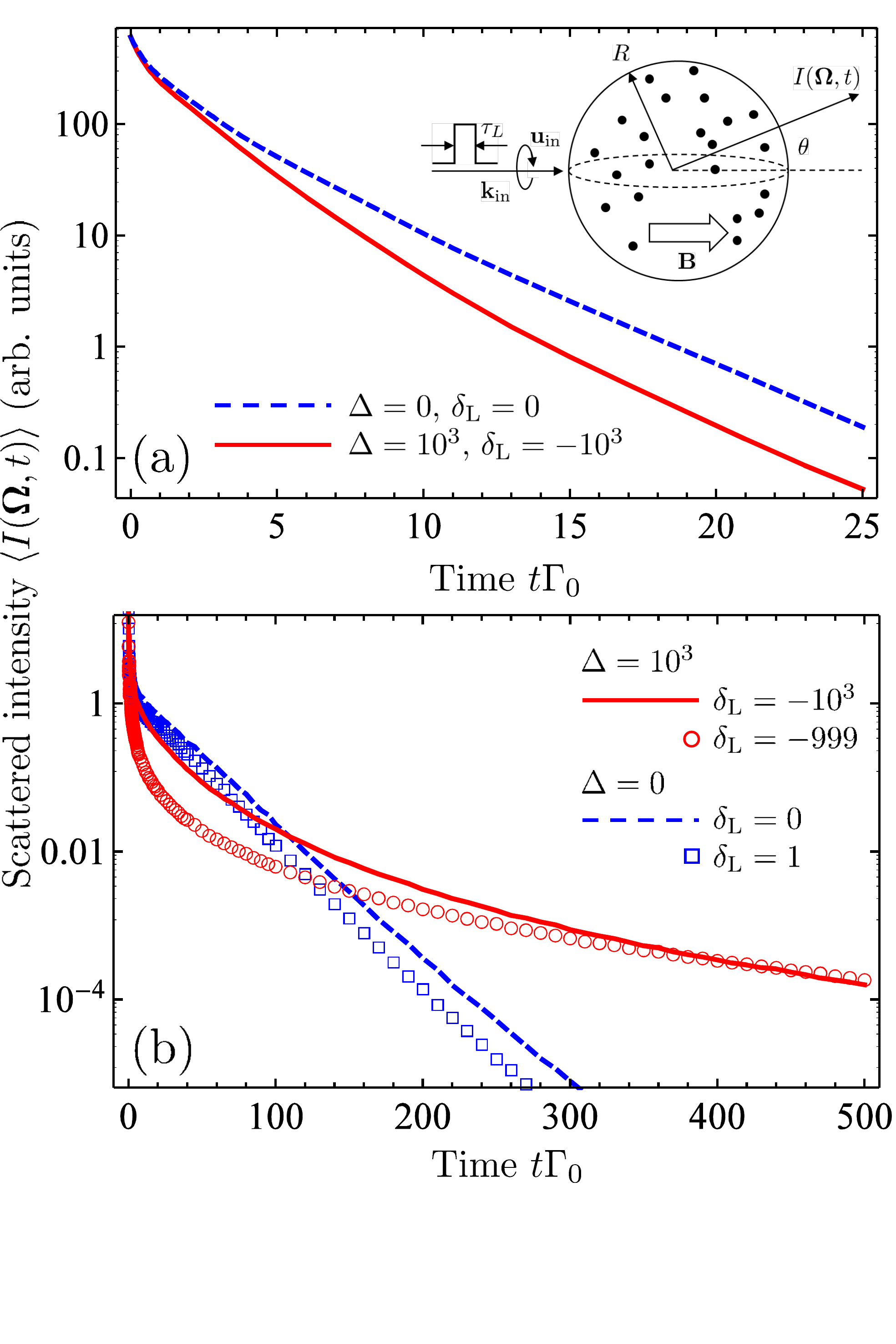}
\vspace*{-15mm}
\caption{Time-dependent fluorescence of a dilute [$\rho/k_0^3 = 1.6 \times 10^{-3}$, panel (a)] and dense [$\rho/k_0^3 = 0.2$, panel (b)] clouds of two-level atoms, with ($\Delta = 10^3$) and without ($\Delta = 0$) magnetic filed. The calculations were performed for the fluorescence in the direction $\theta = \pi/6$ for the left-handed circular polarized rectangular exciting pulse of duration $\tau_{\mathrm{L}} = 5 \Gamma_0^{-1}$ detuned by $\delta_{\mathrm{L}}$ from the fundamental atomic resonance, and for the total number of atoms $N = 1448$ (a) or 838 (b). $t = 0$ corresponds to the end of the exciting pulse. The inset of panel (a) shows a sketch of the considered experimental situation.}
\label{fig_fluo_1}
\end{figure}

When the number density of atoms $\rho = N/V$ is low, the magnetic field speeds up the fluorescence as we show in Fig.\ \ref{fig_fluo_1}(a) for left-handed circular polarized light exciting the transition between the ground state $|g_i \rangle$ characterized by $J_{\mathrm{g}} = m_{\mathrm{g}} = 0$ and the excited state $|e_{im} \rangle$ with  $J_{\mathrm{e}} = 1$, $m = -1$.
This is due to the decrease of the atomic scattering cross-section in the magnetic field \cite{sigwarth13} and is opposite to what could be expected from previous studies of dilute ensembles of atoms with a degenerate ground state ($J_{\mathrm{g}} > 0$) \cite{sigwarth04}. In the latter case, the magnetic field enhances interference effects and therefore should strengthen light trapping and slow down the fluorescence.

In contrast to the limit of low density, at high $\rho$ a dramatic slowing down of fluorescence is observed at long times for detunings $\delta_{\mathrm{L}} \simeq -\Delta$ [see Fig.\ \ref{fig_fluo_1}(b)]. We explain this result by the appearance of spatially localized collective atomic states at large densities $\rho/k_0^3 \gtrsim 0.1$ and strong magnetic fields \cite{skip15}. These localized states have eigenfrequencies $\omega \simeq \omega_0 \pm \Delta \Gamma_0 + \Gamma_0$ and very long lifetimes decreasing exponentially with the sample size $R$ \cite{skip15}. They dominate $\langle I(\mathbf{\Omega}, t) \rangle$ at long times, when all other states excited by the incident beam have died out. Analysis of Fig.\ \ref{fig_fluo_1} allows us to conclude that the magnetic field impacts both far- and near-field properties of inter-atomic interactions, but its influence on the near-field effects (that suppress the disorder-induced localization in the absence of the field \cite{skip14}) is more pronounced. It is worthwhile to note that results similar to those presented in Fig.\ \ref{fig_fluo_1} for $\theta = \pi/6$ and $m = -1$ were also obtained for other scattering angles $\theta$ as well as for right-handed circular polarized incident light that excites transitions between $|g_i \rangle$ and $|e_{im} \rangle$ with $m = 1$. In the latter case, the slowing down of fluorescence is observed for $\delta _{\mathrm{L}} \simeq \Delta$ instead of $\delta _{\mathrm{L}} \simeq -\Delta$ for $m = -1$ in Fig.\ \ref{fig_fluo_1}(b).

\begin{figure}
%% \hspace*{-5mm}
\includegraphics[width=0.9\columnwidth]{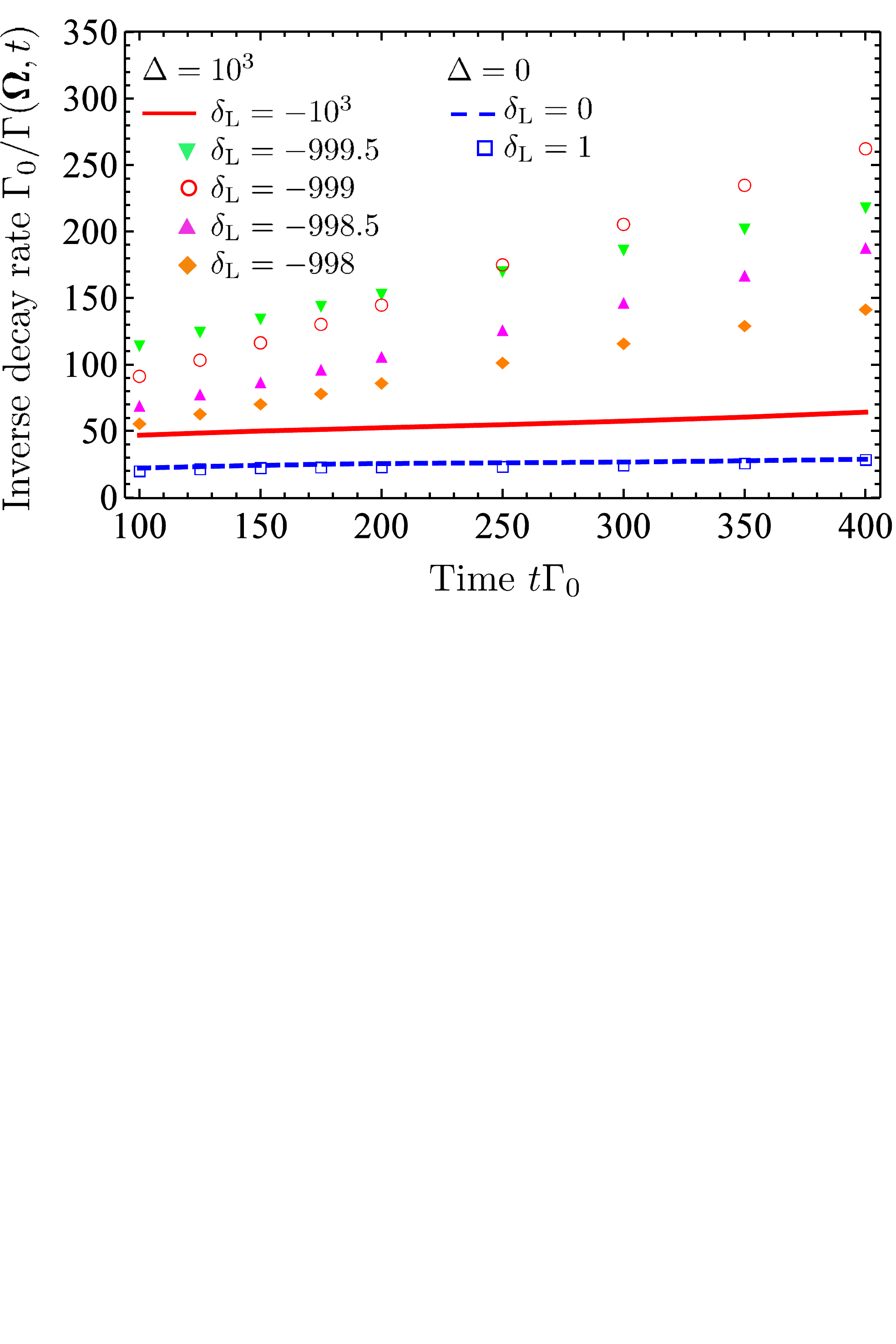}
\vspace*{-6.5cm}
\caption{Inverse fluorescence decay rate $\Gamma$ in a dense atomic cloud in a strong magnetic field ($\Delta = 10^3$) and in the absence of the field ($\Delta = 0$), for different detunings $\delta_{\mathrm{L}}$ of the exciting pulse.
Calculations have been performed for the same parameters as in Fig.\ \ref{fig_fluo_1}(b) but for a much longer exciting pulse ($\tau_{\mathrm{L}} = 10^3 \Gamma_0^{-1}$) to restrict the excitation to a narrow spectral region.}
\label{fig_decay_rate}
\end{figure}

The magnetic field does not only slow down the fluorescence but also modifies the functional form of $\langle I(\mathbf{\Omega}, t) \rangle$ making it nonexponential. Figure \ref{fig_decay_rate} shows the inverse of the decay rate $\Gamma(\mathbf{\Omega}, t) = \partial \ln \langle I(\mathbf{\Omega}, t) \rangle/\partial t$ as a function of time $t$ for $\mathbf{\Omega} = \{ \pi/6, \varphi \}$. We see that in the absence of the field ($\Delta = 0$) the decay rate $\Gamma$ is almost independent of time. This is perfectly consistent with the diffuse nature of radiation transport that we have established under such conditions previously \cite{skip14, fofanov13}. In contrast, in a strong magnetic field ($\Delta = 10^3$), $1/\Gamma$ grows roughly linearly with time which is at odds with the diffusive picture of scattering. The growth rate of $1/\Gamma$ strongly depends on the detuning  $\delta_{\mathrm{L}}$ of the exciting radiation from the atomic resonance and reaches a maximum for $\delta_{\mathrm{L}} \simeq m \Delta + 1$. It is precisely the regime in which localized states are expected to appear in the atomic system \cite{skip15}, which once again confirms the major role of these states for the slow decay of fluorescence reported in Fig.\ \ref{fig_fluo_1}(b). As follows from Fig.\ \ref{fig_decay_rate}, the magnetic field suppresses the fluorescence decay rate by at least an order of magnitude in an atomic cloud that contains only as much as $N \sim 10^3$ atoms. A much stronger effect is expected in larger clouds.

\begin{figure}
%% \hspace*{-5mm}
\includegraphics[width=0.9\columnwidth]{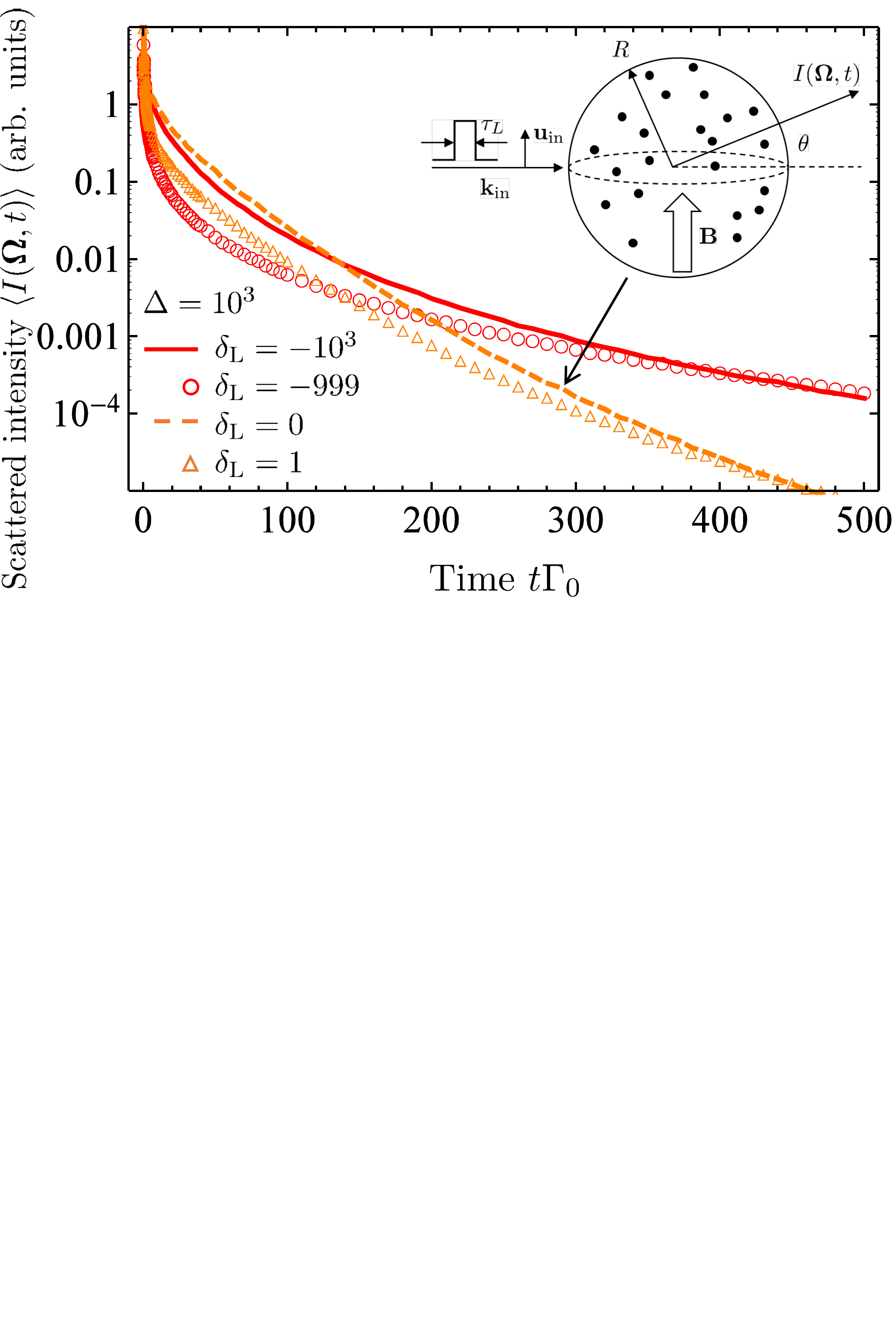}
\vspace*{-6.5cm}
\caption{Time-dependent fluorescence of a dense cloud of two-level atoms ($\rho/k_0^3 = 0.2$) in a strong magnetic field under the same conditions as in Fig.\ \ref{fig_fluo_1}(b) but for excitations of transitions with different magnetic quantum numbers $m$ of the excited state: $m = -1$ [red solid line and circles, the same data as in Fig.\ \ref{fig_fluo_1}(b)] and $m = 0$ (orange dashed line and triangles). The inset shows a sketch of experimental situation for the excitation of $m = 0$ transition.}
\label{fig_fluo_2}
\end{figure}

Because localized states are expected to appear only in the vicinity of resonant frequencies $\omega = \omega_0 \pm \Delta \Gamma_0$ corresponding to transitions from the ground state to the excited states with $m = \pm 1$, but not in the vicinity of  $\omega = \omega_0$ corresponding to the transition to the state with $m = 0$ \cite{skip15}, it is interesting to compare the time-dependent fluorescence for the exciting radiation detuned by $\delta_{\mathrm{L}} \simeq \pm \Delta$ from the atomic resonance with that for $\delta_{\mathrm{L}} \simeq 0$. Such a comparison shown in Fig.\ \ref{fig_fluo_2} reveals that the fluorescence decays much slower for  $\delta_{\mathrm{L}} \simeq -\Delta$ than for $\delta_{\mathrm{L}} \simeq 0$. An experimental observation of such an important difference in decay rates may serve as an evidence of the presence of localized states in the atomic system.

\section{Short- and intermediate-time fluorescence}
\label{sec:short}

\begin{figure}
%% \hspace*{-5mm}
\includegraphics[width=0.9\columnwidth]{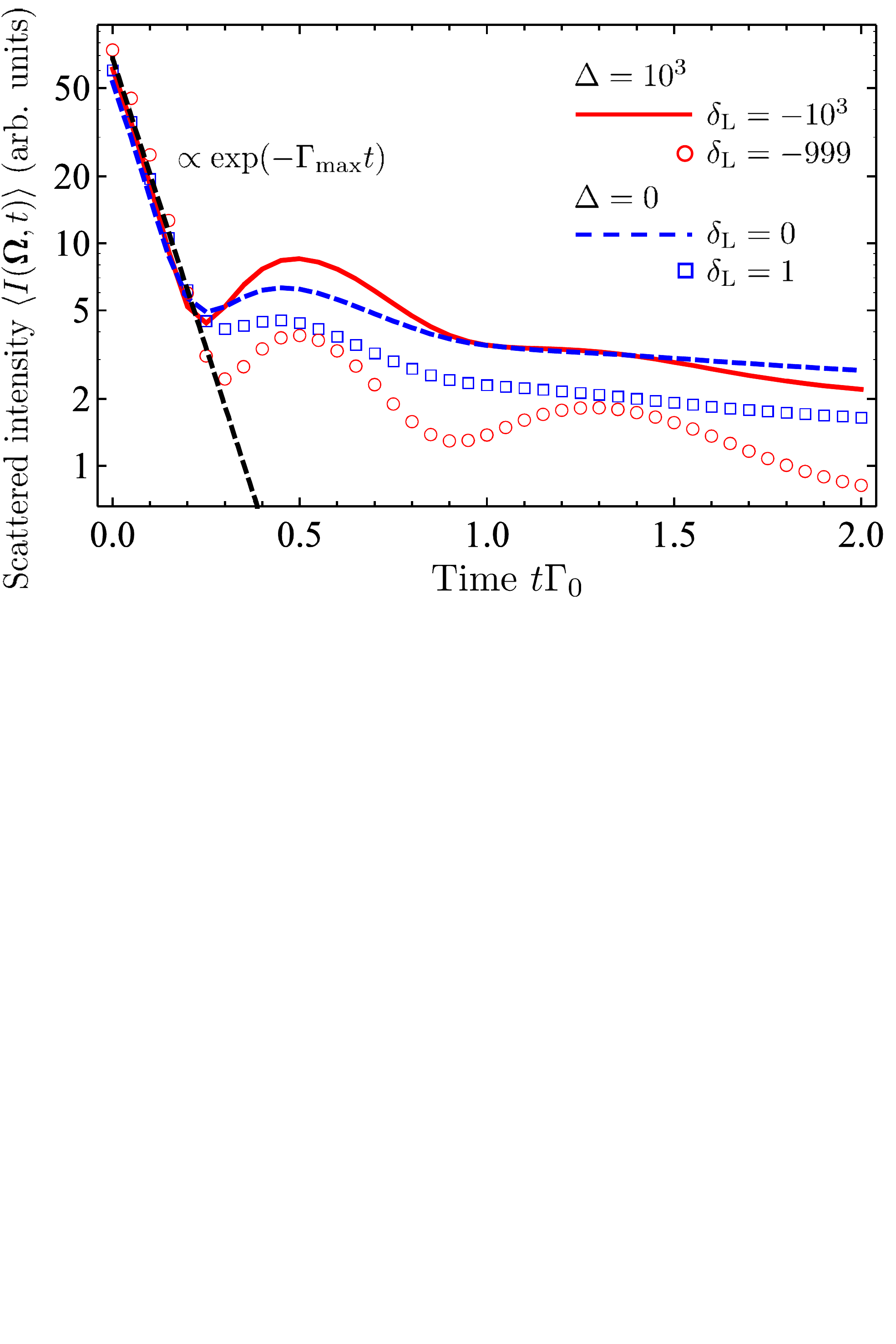}
\vspace*{-6.5cm}
\caption{Short-time dynamics of fluorescence under the same conditions as in Fig.\ \ref{fig_fluo_1}(b). The dashed straight line shows the fast initial decay due to the superradiance phenomenon, with the decay rate $\Gamma_{\mathrm{max}} \simeq 12 \Gamma_0$ following from Eq.\ (\ref{gamma_max}).}
\label{fig_fluo_3}
\end{figure}

In contrast to the long-time decay, the behavior of $\langle I(\mathbf{\Omega}, t) \rangle$ at short time scales is not expected to be sensitive to the presence of localized states. However, this behavior still exhibits a number of interesting features including the phenomenon of Dicke superradiance \cite{dicke54, gross82}. The latter is manifest in a rapid initial decay of  $\langle I(\mathbf{\Omega}, t) \rangle$ (for $t \lesssim 0.2 \Gamma_0^{-1}$ in Fig.\ \ref{fig_fluo_3}) and is hardly sensitive to the frequency $\omega_{\mathrm{L}}$ of the exciting pulse, at least as long as  $\omega_{\mathrm{L}}$ remains close to the resonant frequency of one of the three resonant atomic transitions ($m = 0$, $\pm 1$; only the results for $m = -1$ are shown in Fig.\ \ref{fig_fluo_3}). This is not surprising because the initial decay of $\langle I(\mathbf{\Omega}, t) \rangle$ is dominated by quasi-modes with large decay rates $\Gamma$ that are excited with very similar weights independent from the frequency of incident beam and from the magnetic field. To estimate the decay rate $\Gamma_{\mathrm{max}}$ of the most rapidly decaying quasi-modes, we use an equation for the boundary of the eigenvalue domain of the Green's matrix $G$ obtained in the diffusion approximation \cite{goetschy11, goetschy11c}:
\begin{eqnarray}
|\Lambda - i|^2 = \frac{\sqrt{3} b_0}{2 \pi} \sqrt{\mathrm{Im} \Lambda} \left( 1 + \frac{|\Lambda - i|^2}{|\Lambda - i|^2 + \frac{3}{4} b_0} \right),
\label{gamma_max}
\end{eqnarray}
where $b_0 = 2R/\ell_0$ and $\ell_0 = k_0^2/6\pi\rho$ is the on-resonance mean-free path in the independent-scattering approximation \cite{note1}. Under conditions of Fig.\ \ref{fig_fluo_3}, $b_0 \simeq 75$ and the solution of Eq.\ (\ref{gamma_max}) with $\Lambda = i \Gamma_{\mathrm{max}}/\Gamma_0$ yields $\Gamma_{\mathrm{max}} \simeq 12 \Gamma_0$. This result shown by a dashed line in Fig.\ \ref{fig_fluo_3} describes quite well the initial fast decay of scattered intensity. It is worthnoting, however, that the short-time decay of $\langle I(\mathbf{\Omega}, t) \rangle$ depends on $\mathbf{\Omega}$, with the fastest decay taking place in forward scattering ($\theta = 0$). Thus, $\Gamma_{\mathrm{max}}$ is a good estimation but not an exact result. On the other hand, the fast superradiant decay of $\langle I(\mathbf{\Omega}, t) \rangle$ is seen at large scattering angles (at least up to $\theta \simeq \pi/4$) which distinguishes it from coherent transients emitted in the forward direction by a dilute atomic cloud after an abrupt switch-off of the incident laser \cite{allen87,chalony11}. At longer times $\Gamma_{\mathrm{max}}^{-1} \ll t \lesssim \Gamma_0^{-1}$, the scattered intensity exhibits well pronounced oscillatory behavior that survives ensemble averaging. Even though these oscillations are already visible in the absence of external magnetic field, they are largely amplified by the latter.

\begin{figure}
%% \hspace*{-5mm}
\includegraphics[width=0.9\columnwidth]{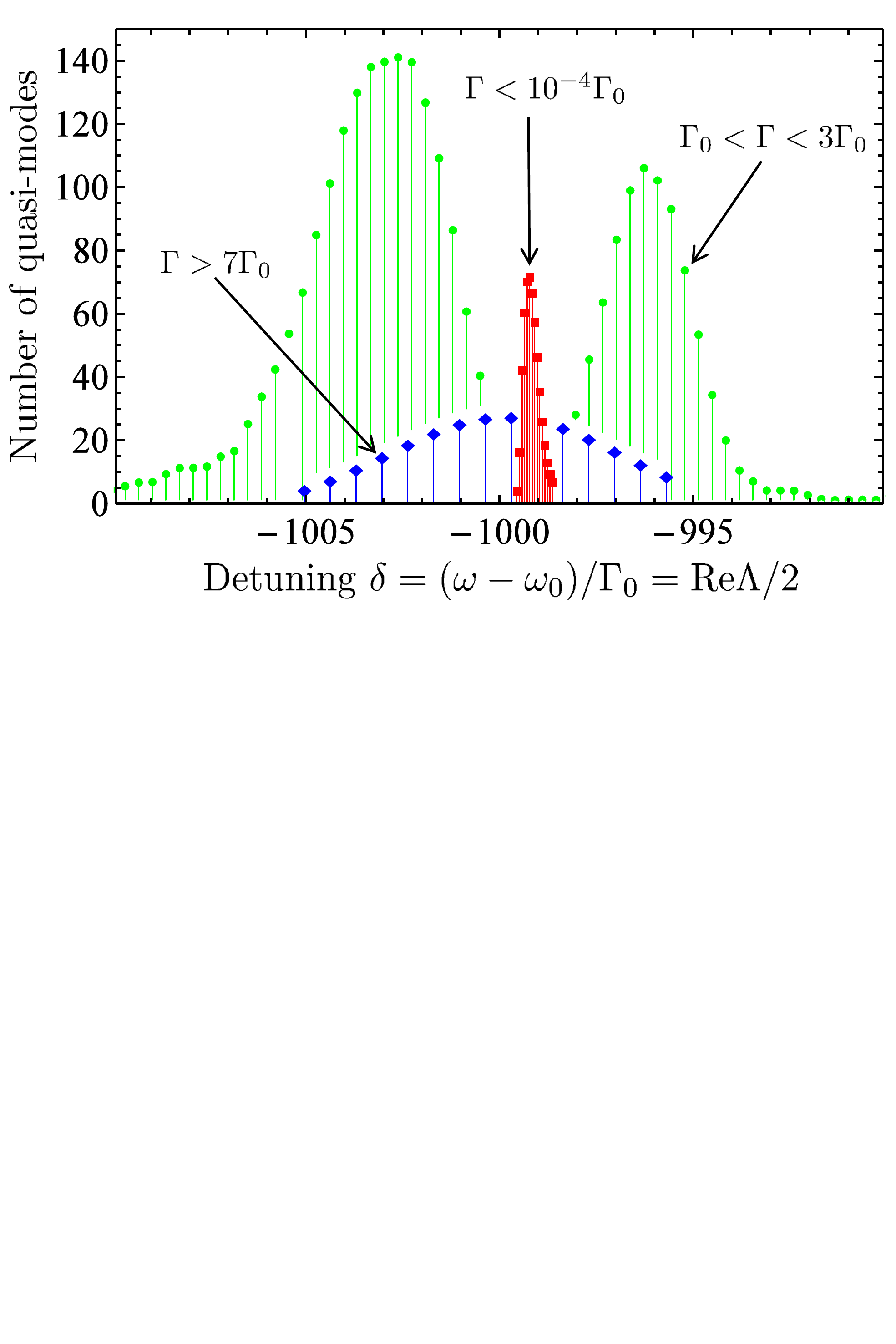}
\vspace*{-6.5cm}
\caption{Spectral distributions of eigenstates with different decay rates $\Gamma$ for a dense ($\rho/k_0^3 = 0.2$) spherical cloud of $N = 4 \times 10^3$ atoms in a strong magnetic field ($\Delta = 10^3$). Only a part of frequency range around $\delta = -\Delta$ is shown; a similar picture is obtained around $\delta = \Delta$.}
\label{fig_states}
\end{figure}

\section{Spectral distribution of quasi-modes}
\label{sec:modes}

A qualitative understanding of different regimes of fluorescence reported above can be achieved by analyzing the spectral distribution of quasi-modes of the system with decay rates $\Gamma = \Gamma_0 \mathrm{Im} \Lambda$ in different characteristic ranges (see Fig.\ \ref{fig_states}). Quasi-modes with large decay rates $\Gamma > 7 \Gamma_0$ can be found in a wide spectral range around the resonant detuning $\delta = - \Delta$ (and also around $\delta = \Delta$, not shown in Fig.\ \ref{fig_states}). These states are responsible for the rapid initial decay of intensity in Fig.\ \ref{fig_fluo_3}. Their wide spectral distribution explains the independence of the initial decay from the detuning $\delta_{\mathrm{L}}$ of the exiting pulse. Quasi-modes with intermediate decay rates $\Gamma_0 < \Gamma < 3 \Gamma_0$ follow peculiar double-peak distributions with minima in the vicinity of $\delta = \pm \Delta$. Quantum beatings between states of this group are at the origin of intensity oscillations in Fig.\ \ref{fig_fluo_3}. Finally, states with extremely low decay rates $\Gamma < 10^{-4} \Gamma_0$ are concentrated around $\delta \simeq \pm \Delta + 1$ and are spatially localized according to our previous analysis \cite{skip15}. These states produce the slow decay of scattered intensity at long times shown in Figs.\ \ref{fig_fluo_1}(b) and \ref{fig_fluo_2}.

\section{Proposal for an experiment}
\label{sec:exp}

Let us now discuss a possible experimental realization of the control scheme proposed above. Although typical ranges of decay times $t < 500/\Gamma_0$ analyzed above correspond to those achieved in recent experiments with cold Rb atoms \cite{guerin16}, the latter are not the best candidates for observation of effects that we report because they do not provide an easily manageable $J_{\mathrm{g}} = 0$ $\to$ $J_{\mathrm{e}} = 1$ optical transition. Examples of atoms that do provide such a transition and, in addition, can be conveniently cooled to low temperatures at large atomic number densities are strontium (Sr) \cite{bidel02,xu03} and ytterbium (Yb) \cite{kuwamoto99,takasu03}. Here we discuss them as potential candidates for experimental realization of phenomena discussed in the Secs.\ \ref{sec:long} and \ref{sec:short}.

\subsection{Energy level structure}

\begin{figure}[!h]
\includegraphics[width=0.85\columnwidth]{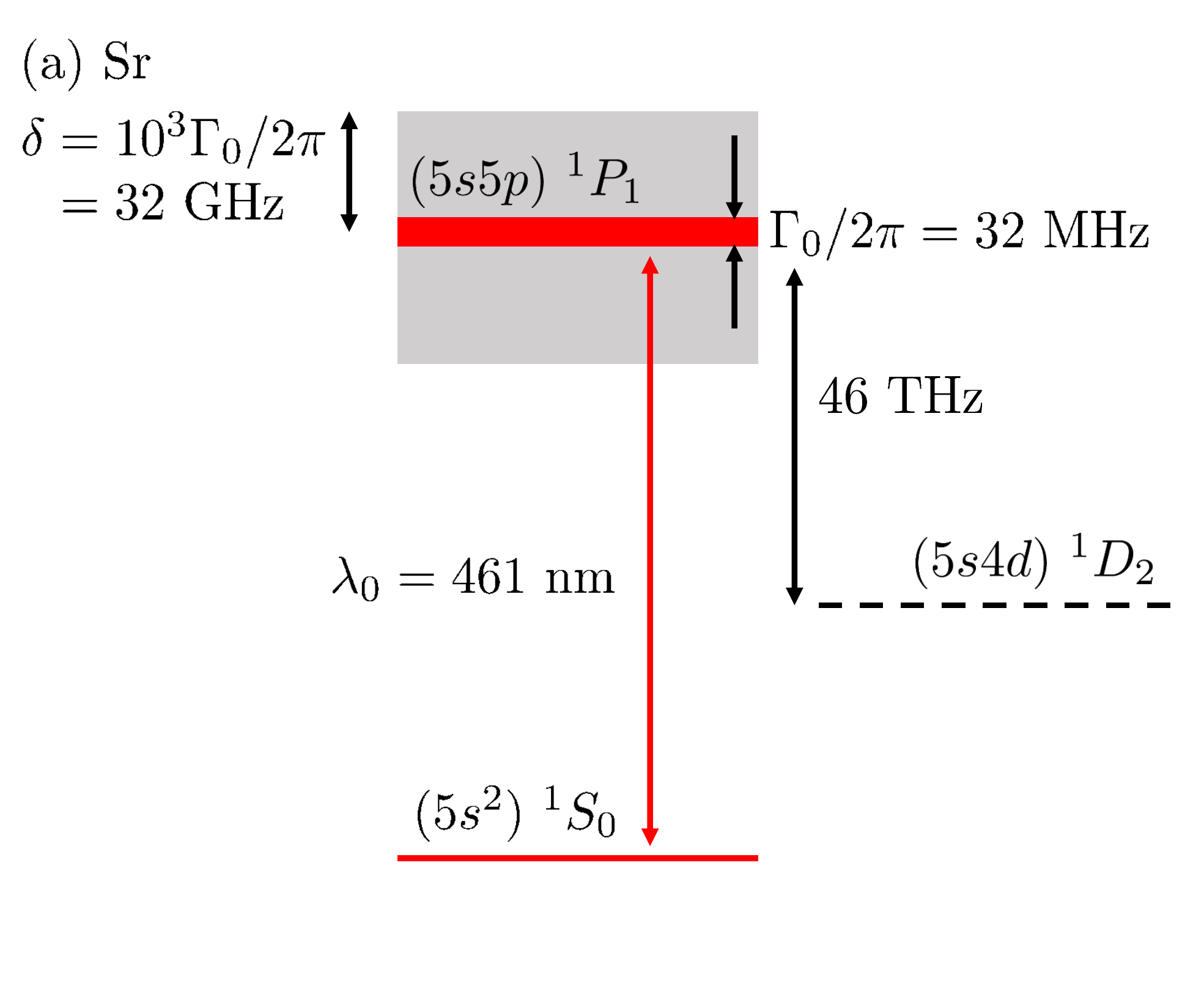}\\
\vspace*{-2mm}
\includegraphics[width=0.85\columnwidth]{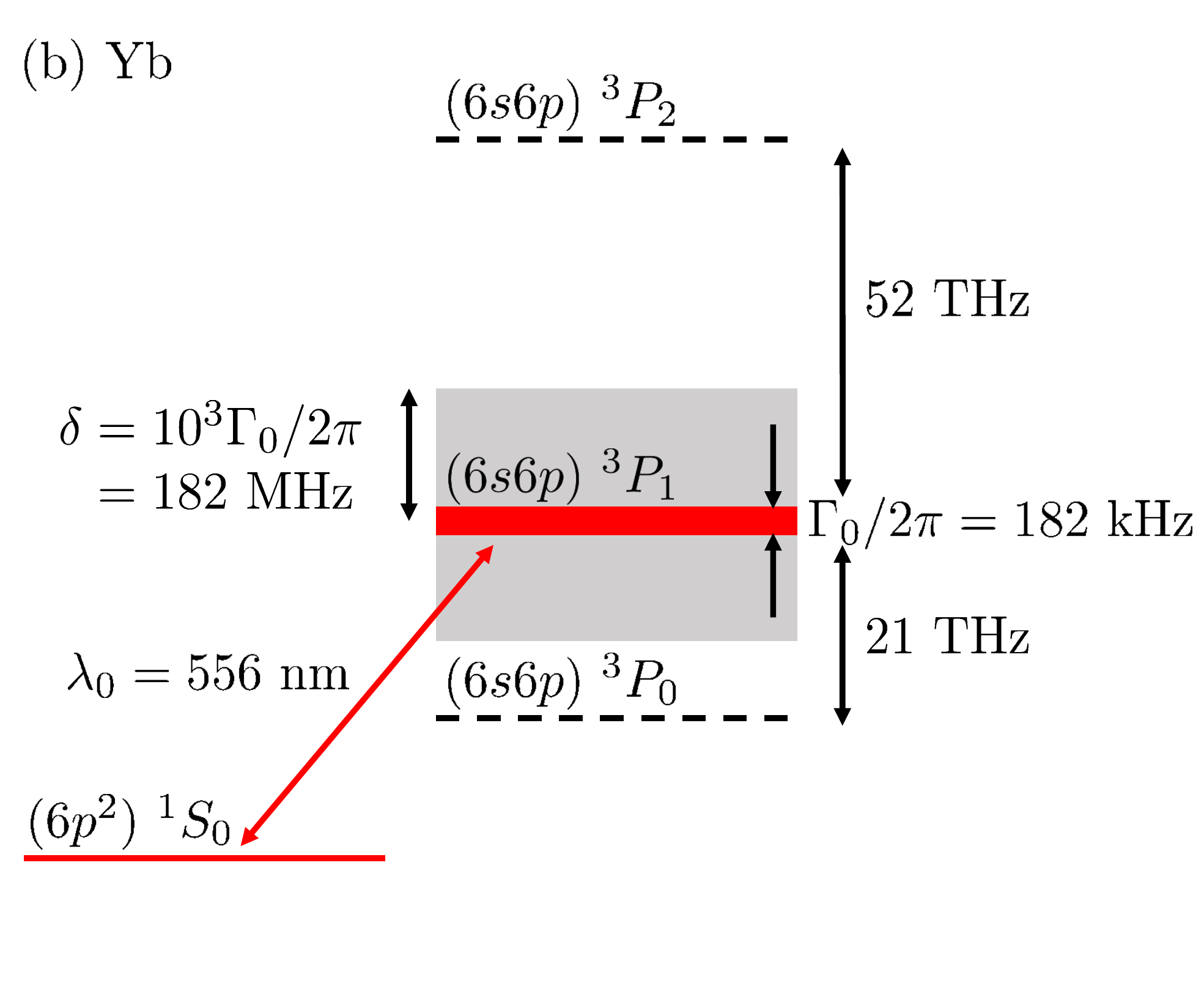}
\vspace*{-7mm}
\caption{Energy-level diagrams of Sr (a) and Yb (b) where we show the two relevant levels (solid red lines) and the levels that are the closest to the excited ones (dashed lines). The large thicknesses of the excited levels symbolize their natural widths $\Gamma_0$ whereas the shaded gray areas show the energy ranges explored by the excited levels in a magnetic field for Zeeman shifts up to $\delta = 10^3 \Gamma_0/2\pi$ (distances between levels are not to scale).}
\label{ed}
\end{figure}

Both strontium and ytterbium have several isotopes with zero nuclear spin $I = 0$. One can choose, e.g., $^{88}$Sr and $^{174}$Yb that are naturally most abundant. Choosing isotopes with $I = 0$ simplifies the analysis because of the absence of the hyperfine level structure. Furthermore, both Sr and Yb have (nearly) closed optical transitions with the required quantum numbers $J_{\mathrm{g}} = 0$ and $J_{\mathrm{e}} = 1$. We will consider the $^1S_0$ $\to$ $^1P_1$ transition resonant with light at a wavelength $\lambda_0 = 461$ nm for Sr and the $^1S_0$ $\to$ $^3P_1$ transition resonant with light at a wavelength $\lambda_0 = 556$ nm for Yb. The relevant energy-level diagrams are shown in Fig.\ \ref{ed} (see also Refs.\ \cite{xu03} and \cite{kuwamoto99}). In addition to the levels corresponding to the ground and excited states of the relevant transitions we also show the energy levels closest to the excited state. The excited state of Sr can decay to the $^1D_2$ state instead of the $^1S_0$ one, but the probability of this is low ($\sim 10^{-5}$) \cite{xu03} and we will neglect this possibility in the following.

\begin{table*}[!ht]
\caption{\label{tab1} Summary of parameters relevant in an experiment. The scattering orders $n_{\mathrm{max}}^{\mathrm{Doppler}}$ and $n_{\mathrm{max}}^{\mathrm{recoil}}$ give the maximum times $t_{\mathrm{max}}^{\mathrm{Doppler,\; recoil}} = n_{\mathrm{max}}^{\mathrm{Doppler,\; recoil}}/\Gamma_0$ up to which Doppler broadening and the atomic recoil effect, respectively, can be neglected assuming that the atomic sample is cooled down to (but not below) its quantum degeneracy temperature $T_c$.}
\begin{tabular}{l|l|l}
~&~$^{88}$Sr~&~$^{174}$Yb~\\ [0.3ex]
\colrule
Transition & $(5s^2)$ $^1S_0$ $\to$ $(5s5p)$ $^1P_1$ & $(6p^2)$ $^1S_0$ $\to$ $(6s6p)$ $^3P_1$ \\[0.3ex]
Wavelength $\lambda_0$, nm & 461 & 556  \\[0.3ex]
Natural linewidth $\Gamma_0/2\pi$, MHz & 32 & 0.182 \\[0.3ex]
Distance from the excited to the nearest other level, THz & 46 & 21 \\[0.3ex]
Magnetic field needed to obtain $\Delta \sim 10^3$, T & 1 & $10^{-2}$ \\[0.3ex]
Zeeman shift $\delta = 10^3 \Gamma_0/2\pi$, GHz & 32 & 0.182 \\[0.3ex]
Scattering order $n_{\mathrm{max}}^{\mathrm{Doppler}}$ at $\rho/k_0^3 = 1.6 \times 10^{-3}$ & 530 & 34  \\[0.3ex]
Scattering order $n_{\mathrm{max}}^{\mathrm{Doppler}}$ at $\rho/k_0^3 = 0.2$ & 180 & 12  \\[0.3ex]
Scattering order $n_{\mathrm{max}}^{\mathrm{recoil}}$ & 1600 & 25 \\[0.3ex]
\end{tabular}
\end{table*}

As follows from the figures, the energy-level structures of both Sr and Yb are compatible with the approximation of a two-level atom because even in large magnetic fields levels do not mix and the considered transitions remain well decoupled from the other ones. The maximum Zeeman shift $\delta \simeq 32$ GHz (182 MHz) of the excited level is 3 (5) orders of magnitude smaller than the distance $46$ THz ($21$ THz) from the latter to the nearest neighboring energy level for Sr (Yb). In addition, high densities of cold Yb atoms have been reached in experiments, up to $\rho = 4.7 \times 10^{14}$ cm$^{-3}$ \cite{takasu03}. This corresponds to $\rho/k_0^3 \simeq 0.3$ and is even larger than $\rho/k_0^3 = 0.2$ that we assume in Fig.\ \ref{fig_fluo_1}(b) and Figs.\ \ref{fig_decay_rate}--\ref{fig_states} to illustrate our results. The dimensionless Zeeman shift $\Delta = 2\pi \delta/\Gamma_0 = 10^3$ that we use corresponds to magnetic fields $B \sim 1$ T ($10^{-2}$ T) for Sr (Yb).  $B \sim 10^{-2}$ T is within the reach of state-of-the-art experiments in cold atomic systems (see, e.g., Ref.\ \cite{sigwarth04}) whereas $B = 1$ T may require some efforts to be realized but is anyway achievable without any doubt.

\subsection{Doppler broadening}

An important experimental aspect is the broadening of the spectrum of scattered radiation due to Doppler effect. The role of Doppler broadening in multiple scattering of light by cold atoms has been previously studied in Refs.\ \cite{labeyrie03,labeyrie05,labeyrie06}. Doppler broadening is neglected in our calculation that assumes that the spectrum of light does not change upon propagation in the atomic cloud. The residual 1D rms thermal velocity $v$ of an atom in a cloud at a temperature $T$ can be estimated from a relation $Mv^2 = k_{\mathrm{B}}T$, where $M$ is the atomic mass and $k_{\mathrm{B}}$ is the Boltzmann constant. The magnitude of the Doppler shift per scattering event is of order $k_0 v$ and should remain less than the natural linewidth $\Gamma_0$. $k_0 v \ll \Gamma_0$ implies $T \ll M \Gamma_0^2/k_0^2 k_{\mathrm{B}} \approx 2$ K for Sr and 200 $\mu$K for Yb atoms. This condition can be easily obeyed not only for Sr but also for Yb which has been cooled to at least 20 $\mu$K \cite{kuwamoto99}. In the multiple scattering regime, a photon acquires a random frequency shift of order $k_0 v$ at each scattering and its frequency performs a random walk in the frequency space. The number $n$ of scattering events necessary for the spectrum of a photon to reach a width of order $\Gamma_0$ is then $n \sim (\Gamma_0/k_0 v)^2$.
An accurate calculation taking into account the distribution of atomic velocities yields a quite stringent condition for the critical scattering order  $n_c$ up to which one can neglect the thermal motion of atoms: $n_c = 3^{1/3}(\Gamma_0/k_0 v)^{2/3}$ \cite{labeyrie06}. Note that the horizontal axes of Figs.\ 1--4 can be interpreted as the scattering order $n = t\Gamma_0$ because each scattering event takes a time of order $1/\Gamma_0$.

We see from the above that whereas the short- and intermediate-time fluorescence $n \lesssim 2 < n_c$ (see Fig.\ \ref{fig_fluo_3}) can be readily made insensitive to Doppler broadening, suppressing the impact of the latter in the long-time regime $n \sim 10^2 < n_c$ (as in Figs.\ \ref{fig_fluo_1}--\ref{fig_fluo_2}) is challenging and requires cooling atoms below
\begin{eqnarray}
T_{\mathrm{max}} = \frac{3 \Gamma_0^2 M}{k_0^2 k_{\mathrm{B}} n^3}.
\label{tmax}
\end{eqnarray}
$T_{\mathrm{max}} \simeq 7$ $\mu$K for Sr and 0.6 nK for Yb. On the other hand, the lowest temperature to which an atomic gas can be cooled without entering the quantum degeneracy regime is \cite{pitaevskii03}:
\begin{eqnarray}
T_c = 3.31 \frac{\hbar^2 \rho^{2/3}}{k_{\mathrm{B}} M}.
\label{tc}
\end{eqnarray}
It follows from Eq.\ (\ref{tmax}) that at this temperature, the Doppler broadening can be neglected for scattering orders $n = t \Gamma_0 < n_{\mathrm{max}} = t_{\mathrm{max}} \Gamma_0 \simeq (\Gamma_0 M/\hbar k_0 \rho^{1/3})^{2/3}$.
We find $n_{\mathrm{max}} \simeq 34$ (12) for Yb and $n_{\mathrm{max}} \simeq 530$ (180) for Sr at a density $\rho/k_0^3 = 1.6 \times 10^{-3}$ (0.2) that we assumed in Fig.\ \ref{fig_fluo_1}(a) (Fig.\ \ref{fig_fluo_1}(b) and Figs.\ \ref{fig_fluo_2}--\ref{fig_states}). It is also important to note that even for $n > n_{\mathrm{max}}$, the phenomena that we discuss do not disappear altogether but simply become less pronounced. The widening of the spectrum can be taken into account in the calculation to describe such a situation, but this falls beyond the scope of the present work.

\subsection{Atomic recoil}

Atomic recoil is yet another phenomenon modifying the spectrum of scattered light. A photon transfers a mechanical momentum of order $\hbar k_0$ to the atom on which it is scattered. The atom then acquires a velocity $v \sim \hbar k_0/M$ that eventually plays a role similar to the thermal velocity. As in the estimation of the Doppler effect, we need $k_0 v \ll \Gamma_0$. This condition is well verified for both Sr ($k_0 v/2\pi \simeq 20$ kHz $\ll \Gamma_0/2\pi = 32$ MHz) and Yb ($k_0 v/2\pi \simeq 7$ kHz $\ll \Gamma_0/2\pi = 182$ kHz). In contrast to Doppler shift that has a random sign, the frequency shift due to recoil is always negative and the number of scattering events needed for the recoil effect to shift the frequency of incident light by $\Gamma_0$ is $n_{\mathrm{max}} = t\Gamma_0 \sim \Gamma_0/k_0 v \approx 1600$ for Sr and 25 for Yb.

\subsection{Discussion and alternative experiments}

A summary of estimated parameters for Sr and Yb atoms is presented in Table \ref{tab1}. The ensemble of results indicates that dense clouds of Sr atoms are good candidates for the experimental realization of the control over light trapping in a cold atomic gas proposed in this work although cooling to low temperatures ($\sim 1$ $\mu$K) and applying large magnetic fields (up to 1 T) may be necessary. For Yb atoms, our theory can be applied for short and intermediate times but needs to be extended by including Doppler broadening and recoil effects to describe the long-time regime ($n \sim 10^2$).

It is worthwhile to note that the two main difficulties on the way towards experimental realization of the control scheme proposed in the present work---the Doppler broadening of the pulse and the atomic recoil effect---can be avoided or, at least, softened to a considerable extent by using a different, solid-state experimental realization of the Hamiltonian (\ref{ham}) in which atoms are replaced by quantum dots (``artificial atoms'') embedded in a transparent solid matrix or suspended in a viscous liquid. Several experiments exist to date in which the resonant optical response of quantum dots was shown to be similar to that of atoms (see, e.g., Refs.\ \onlinecite{xu12, konth12, ulhaq12}). However, a solid-state system may present other challenges, such as, e.g., more or less important fluctuations of physical properties of $N$ quantum dots which are difficult to fabricate all strictly identical (in contrast to real atoms that are all exactly the same), a potential difficulty of isolating a pair of energy levels of a quantum dot from the rest of the spectrum, mechanical vibrations (phonons), etc. A detailed analysis of the impact of these complications on time-dependent fluorescence of a large ensemble of quantum dots is clearly beyond the scope of the present work.

Finally, our calculation is valid for a transition between ground and excited states with total angular momenta $J_{\mathrm{g}} = 0$ and $J_{\mathrm{e}} = 1$, respectively. An analogous calculation for a transition with larger $J_{\mathrm{g}}$, $J_{\mathrm{e}}$ (like, e.g., the transitions of Rb atoms widely used in cold-atom experiments \cite{guerin16,labeyrie03,labeyrie05,labeyrie06}) would be much more involved. However, a strong magnetic field lifts the degeneracy of both ground and excited states and thus decouples transitions corresponding to different magnetic quantum numbers $m_{\mathrm{g}}$ and $m_{\mathrm{e}}$. A nondegenerate transition with given $m_{\mathrm{g}}$, $m_{\mathrm{e}}$ is not equivalent to a simple two-level system in the absence of the field and transitions with different $m_{\mathrm{g}}$, $m_{\mathrm{e}}$ may exhibit very different properties in regard to spatial localization of collective quasi-modes \cite{skip15}. Even though the solution of the full problem is beyond the scope of this work, it is not impossible that some of these transitions may behave similarly to $J_{\mathrm{g}} = 0$, $m_{\mathrm{g}} = 0$ $\to$ $J_{\mathrm{e}} = 1$, $m_{\mathrm{g}} = \pm 1$ transition, leading to the appearance of spatially localized eigenstates at high densities of atoms and allowing for an efficient control of fluorescence by a magnetic field. The corresponding experiment would be facilitated by the strong expertise of many experimental groups in cooling and controlling clouds of Rb atoms as opposed to relatively scarce experiments with Sr atoms.

\section{Conclusion}
\label{sec:concl}

In conclusion, we predict that an external magnetic field should allow for an efficient control of light trapping in large ensembles of cold atoms. We illustrate this result by considering the limit of a very strong field but similar conclusions hold in weaker fields as well.
Most of our calculations were performed for a moderate number of atoms $N \sim 10^3$. However, previous work has shown that distributions of quasi-mode frequencies and lifetimes as well as the spatial localization properties of quasi-modes---the key physical quantities behind our main results---are similar for larger systems (at least up to $N \sim 10^4$) \cite{skip15,skip14}. We therefore believe that our conclusions apply to larger systems as well.
Our theoretical results can be verified in experiments and constitute an important step towards creation of controllable optical elements for future photonic devices to be used for quantum information storage and processing. In addition, they suggest a practical way of observing the disorder-induced localization of light expected to take place in dense atomic ensembles in strong magnetic fields \cite{skip15}.
Recent studies have shown that the search for localization of light using time-resolved experiments is delicate because measured signals can contain contributions due to other, irrelevant phenomena (nonlinear effects \cite{sperling14} or fluorescence \cite{sperling16} in dielectric media). In atomic systems, such phenomena as, e.g., super- and subradiance \cite{dicke54, gross82}, the coherent flash effect \cite{chalony11}, etc. can take place in parallel with the disorder-induced localization but, in contrast to dielectric media, a complete physical model taking into account all these phenomena is available [our Eqs.\ (\ref{ham}) and (\ref{green})].  The present work provides a guide to isolate the localization phenomenon by a proper choice of experimental conditions (frequency detuning of the probe beam, scattering angle, time, magnetic field, polarization, etc.).

\begin{acknowledgments}
We acknowledge financial support from the Agence Nationale de la Recherche (Grant No. ANR-14-CE26-0032 LOVE), the Russian Foundation for Basic Research (Grant No. RFBR-15-02-01013), and the National Science Foundation (Grant No. NSF-PHY-1068159).

\end{acknowledgments}


\begin{thebibliography}{99}
\bibitem{note1}
Note that $\ell_0$ calculated in the independent-scattering approximation is a useful parameter but it is different from the actual mean-free path $\ell(\omega)$ in a dense atomic system where dependent scattering plays an important role. Perturbation of self-consistent analytical methods \cite{vantiggelen90} as well as numerical simulations \cite{fofanov11} can be used to estimate $\ell(\omega)$.
\end{thebibliography}
\end{document}